**$NiGa_2O_4$ interfacial layers in $NiO/Ga_2O_3$ heterojunction diodes at high temperature**

Kingsley Egbo[1,*], Emily M. Garrity[2], William A. Callahan[1,2], Chris Chae[3], Cheng-Wei Lee[2], Brooks Tellekamp[1], Jinwoo Hwang[3], Vladan Stevanovic[2], Andriy Zakutayev[1, †]

[1] National Renewable Energy Laboratory, Golden, CO 80401, USA

[2] Colorado School of Mines, Golden, CO 80401, USA

[3] Department of Materials Science and Engineering, The Ohio State University, Columbus, Ohio 43210, USA

*Kingsley.Egbo@nrel.gov, †Andriy.Zakutayev@nrel.gov

**Abstract:**

$NiO/Ga_2O_3$ heterojunction diodes have attracted attention for high-power applications, but their high temperature performance and reliability remains underexplored. Here we report on the time evolution of the static electrical properties in the widely studied $p$-$NiO$/$n$-$Ga_2O_3$ heterojunction diodes and formation of $NiGa_2O_4$ interfacial layers when operated at 550 °C. Results of our thermal cycling experiment show initial leakage current increase which stabilizes after sustained thermal load, due to reactions at the $NiO$-$Ga_2O_3$ interface. High-resolution TEM microstructure analysis of the devices after thermal cycling indicates that the $NiO$-$Ga_2O_3$ interface forms ternary compounds at high temperatures, and thermodynamic calculations suggest the formation of the spinel $NiGa_2O_4$ layer between $NiO$ and $Ga_2O_3$. First-principles defect calculations find that $NiGa_2O_4$ shows low $p$-type intrinsic doping, and hence can also serve to limit electric field crowding at the interface. Vertical $NiO/Ga_2O_3$ diodes with intentionally grown 5 nm thin spinel-type $NiGa_2O_4$ interfacial layers show excellent device ON/OFF ratio of > $10^{10}(\pm3$ V), $V_{ON}$ of ~1.9 V, and breakdown voltage of ~ 1.2 kV for an initial unoptimized 300 $\mu$m diameter device. These $p$-$n$ heterojunction diodes are promising for high-voltage, high temperature applications.

The promise of electrical functionality at much higher temperatures compared to existing silicon technologies is one of the major drivers of development in several ultrawide bandgap (UWBG) semiconductor based electronic devices[1]. High temperature operation capability is especially desirable in power devices which find applications in several extreme environmental conditions. Among known ultrawide bandgap semiconductors, $β$-$Ga_2O_3$ with a bandgap in the range of 4.5-5.0 eV, and theoretical predicted breakdown electric field of 8 MV/cm, shows remarkable potential for high power, high temperature applications[2]. However, practical development of robust UWBG based power devices capable of reliable operation at high temperatures (>500°C) requires device interface, contact, and interconnect optimization to eliminate thermal induced premature breakdown, leakages, and possible device failure. Increase in junction temperatures due to Joule self-heating at higher power levels also presents another challenge for > 10 A high-power devices, even if operated close to ambient temperature. These considerations are especially important for $β$-$Ga_2O_3$-based devices due to the relatively low thermal conductivity in $β$-$Ga_2O_3$ compared to other UWBG materials[3,4].

Testing of the device characteristics at the desired operating temperature can provide insights on the stability of the electrical properties. $β$-$Ga_2O_3$ heterojunction devices utilizing $p$-type Nickel Oxide (NiO) have been most widely explored due to nickel oxide's favorable band alignment with $β$-$Ga_2O_3$ leading to high energy barrier height. Hence, remarkable room temperature (RT) breakdown voltages and correspondingly high Baliga figures of merit have been reported for NiO/$β$-$Ga_2O_3$ heterojunctions with various thicknesses of $β$-$Ga_2O_3$ drift layer[5–17]. While NiO-$Ga_2O_3$ based devices have shown exceptional breakdown characteristics, their application in high temperature environments has not been widely explored. Previously, our team showed I-V-T characterization on NiO-$Ga_2O_3$ diode with potential for operation up to 400°C with ~ $10^6$ current rectification[18]. Other groups have also shown NiO/$Ga_2O_3$ high breakdown voltage devices capable of operating at 250°C and 275°C respectively[19,20]. However, continuous operation of these NiO-$Ga_2O_3$ devices at high temperatures is yet to be studied.

In this work, we explore the potential of an interfacial $NiGa_2O_4$ layer to enable NiO/$β$-$Ga_2O_3$ vertical heterojunction diodes with robust continuous operation at high temperature. The NiO/$β$-$Ga_2O_3$ diodes measured for >200 hours and 25 thermal cycles up to 550 °C show significant degradation of electrical properties, which from transmission electron microscopy can be attributed to chemical reaction and intermixing at the NiO-$Ga_2O_3$ interface. Thermodynamic calculations suggest a high likelihood for the formation of a ternary compound at this interface – the Ni-Ga-O spinel of the form $AB_2O_4$. Defect calculations suggest that $NiGa_2O_4$ is intrinsically weakly *p*-type with predicted O-rich acceptor carrier density of ~$10^{11}$-$10^{14}$ cm$^{-3}$ at equilibrium conditions, which can help reduce field crowding thus promoting improved device breakdown[21]. To realize this promise, we intentionally grow spinel-type $NiGa_2O_4$ as a 5nm thin interlayer at NiO/$Ga_2O_3$ interface. Fabricated NiO/$NiGa_2O_4$/$Ga_2O_3$ vertical heterojunction diodes on a 5-$\mu$m thick HVPE-$Ga_2O_3$ sample show a rectification ratio of >$10^{10}$ ($\pm 3$ V). Breakdown voltage ($V_{br}$) of 1.2 kV for a large area (300 $\mu$m) device without optimized electric field management techniques is obtained, compared to a $V_{br}$ of 700 for a similar NiO/$Ga_2O_3$ device without the intentionally grown thin $NiGa_2O_4$ layer. This NiO/$NiGa_2O_4$/$Ga_2O_3$ structure is promising for optimizing the device breakdown by spreading the peak electric field, and for passivating interfacial reactions to achieve high voltage devices that can be operated continuously at high temperatures.

The vertical $NiO_x$/$\beta$-$Ga_2O_3$ heterojunction diode sample used for thermal cycling experiment was fabricated on a 1-$\mu$m lightly Si-doped ($3\times10^{16}$ cm$^{-3}$) *n*-type $\beta$-$Ga_2O_3$ drift layer grown on a conductive bulk (001) $\beta$-$Ga_2O_3$ substrate (NCT). The schematic of the device cross-section is shown in Figure 1(a). Ohmic contact of Ti/Au (5 nm/100 nm)[22] was deposited on the backside of the substrate via e-beam evaporation, followed by a rapid thermal annealing (RTA) in $N_2$ ambient at 550 °C for 1 min. Next, the NiO were grown by pulsed laser deposition as described previously[18]. For the front-side ohmic contact on NiO, contact aligner lithography was used to make circular contact patterns of 50-300 $\mu$m diameter followed by e-beam evaporation of Ni/Au (30nm/100nm). To test the device stability under sustained thermal stress, the fabricated NiO/$\beta$-$Ga_2O_3$ vertical heterojunction diodes were subjected to multiple temperature cycle between 25-550°C (up to 25 cycles over >200 hours) with in-operando I-V characterization. Details of the thermal cycling experiment instrumentation were described previously[22].

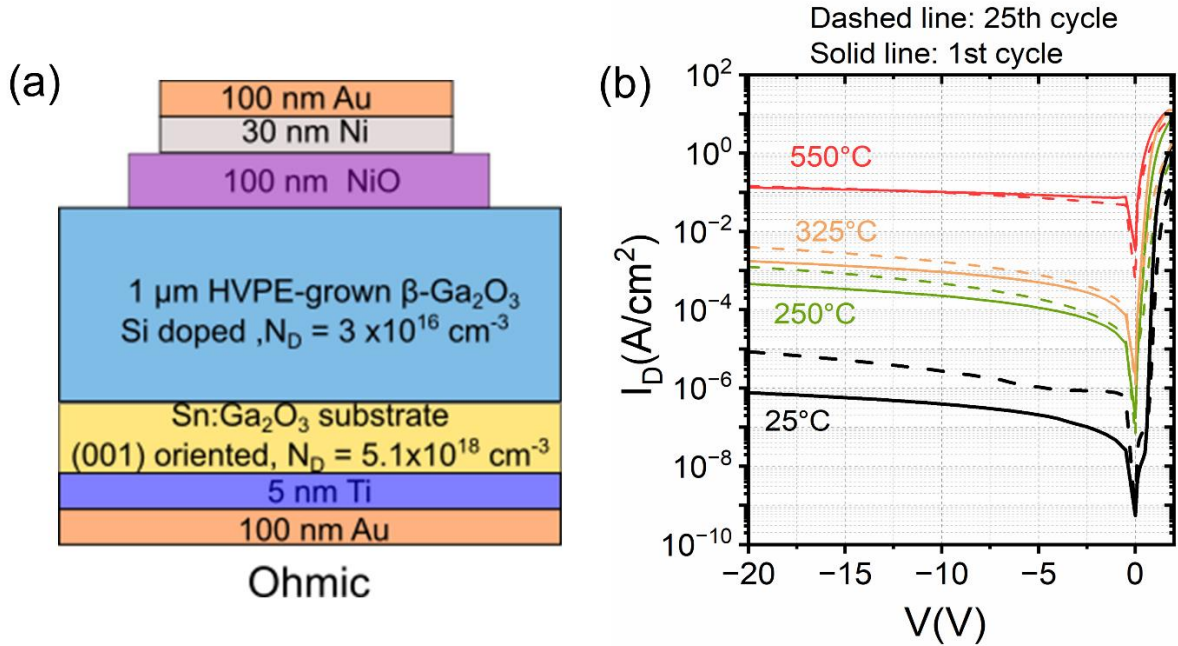

Figure 1: (a) Schematic cross-section of the NiO/$\beta$-Ga$_2$O$_3$ heterojunction diode device stack. (b) J-V curves of the device taken at select operating temperatures before and after the completion of 25 high temperature thermal cycles between 25-550°C. Solid lines are measured at the start of cycle 1 and dashed lines are recorded at the end of cycle 25.

Device J-V characteristics measured during the thermal cycling experiment conducted on a 300$\mu$m diameter NiO/$\beta$-Ga$_2$O$_3$ heterojunction diode are shown in Figure 1(b) before and after 25 thermal cycles conducted between 25°C -550°C. We observe ~ 1 order of magnitude increase in the leakage current density at -20V bias for measurements at 25°C after 25 thermal cycles, but the change is less pronounced for measurements at 550°C. Figure S1 (Supplementary Material) shows the expanded forward bias data from Figure 1(b) of NiO/$\beta$-Ga$_2$O$_3$ heterojunction diode at different cycling temperatures. Similar degradation in the forward current is observed for the different temperatures after the 25 cycles. This differences (increase in leakage current and decrease in forward current) points to the presence of a thermal induced degradation process, likely an interfacial reaction that saturates at 550°C. An indication of this self-limiting temperature-induced interfacial reaction can be seen in the analysis of the leakage current density at -20V for the different temperatures across the 25 cycles as shown in Figure S2 (Supplementary Material). We observe that the increase in the leakage current with cycling present in Figure 1(b) was only up to the 5$^{th}$ cycle. After the 5$^{th}$ cycle the changes in the leakage current became negligible, suggesting the saturation of the interfacial reaction. We focus on the NiO-$Ga_2O_3$ interfacial reaction, because our previous work found that Ti/Au(5nm/100nm) ohmic contacts to $β$-$Ga_2O_3$ show stable electrical properties for operation at 600°C[22].

To explore the NiO/$β$-$Ga_2O_3$ interfacial reaction, cross-section scanning transmission electron microscopy (STEM) was performed on the NiO/$β$-$Ga_2O_3$ device after the thermal cycling experiment. The TEM foil was prepared using the Helios NanoLab 600 DualBeam focused ion beam (FIB), and ion milled at 900V with a Fischione Nanomill to clean the surface of the foil. The STEM investigation was performed on an aberration-corrected Themis-Z from Thermo Fisher Scientific. Figure (2) shows the cross-section of the NiO and $β$-$Ga_2O_3$ interface. Between the NiO and $β$-$Ga_2O_3$ interface is a layer approximately 2nm in thickness where cubic spinel $NiGa_2O_4$ begins to form. The atomic models of $NiGa_2O_4$ along the [110] direction match the atomic structure of the measured $NiGa_2O_4$ region. There appears to be some level of disorder within the $NiGa_2O_4$ layer as the $NiGa_2O_4$ structure is not uniform, showing different possible rotations or orientations of the $NiGa_2O_4$ structure within the 2 nm region of $NiGa_2O_4$. We also note that these layers might be Ni-substituted $γ$-phase $Ga_2O_3$[23] defective spinel.

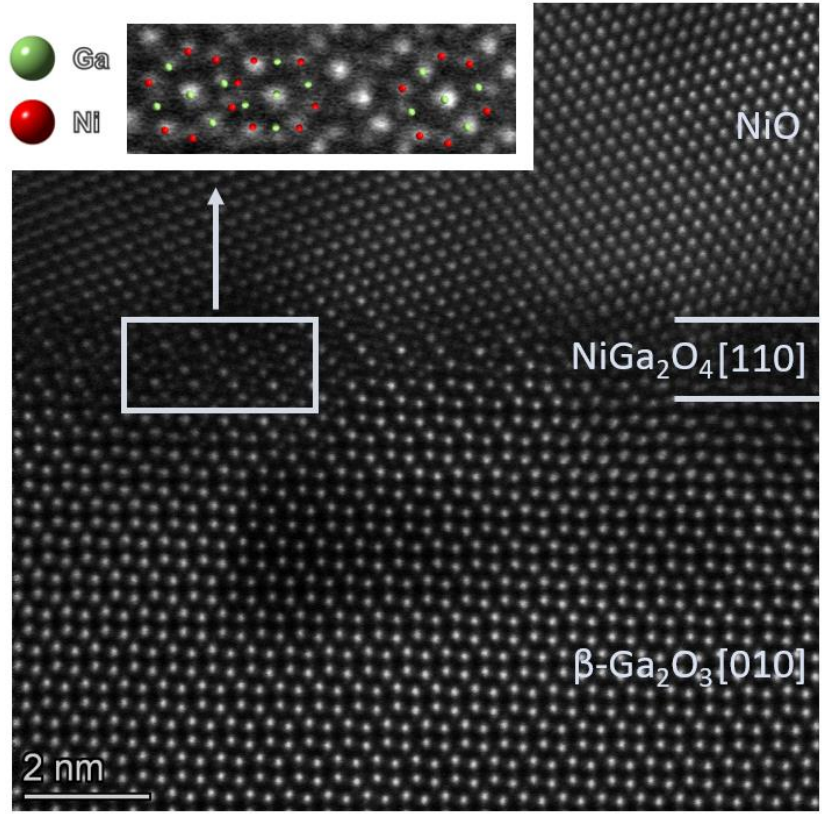

Figure 2: TEM Cross-section image of the NiO/$\beta$-Ga$_2$O$_3$ heterojunction diode after thermal cycling showing the [010] zone for $\beta$-Ga$_2$O$_3$, NiO, and the [110] zone for NiGa$_2$O$_4$. The inset shows a larger view of the NiGa$_2$O$_4$ layer, where the atomic model for [110] NiGa$_2$O$_4$ is overlaid on the STEM image to show the structural match.

To explain the possible outcomes of interfacial reactions in the NiO-Ga$_2$O$_3$ heterostructure during operation at high temperature, as observed in electrical data from the thermal cycling experiment in Figure 1 and the microstructure analysis using TEM in Figure 2, we study the thermodynamic phase diagram of the Ni-Ga-O system. We follow the approach recently developed[24] and calculate the oxygen partial pressure *vs.* temperature ($pO_2$-$T$) phase diagrams using a density functional theory dataset (NREL MatDB[25]) and correction scheme[26,27] that ensures chemical accuracy (~ 50 meV/atom) and captures temperature dependency. Details of the calculations are described in the Supplementary material. Figure 3 shows the calculated phase diagram of Ga-Ni-O system for a wide range of temperatures and oxygen partial pressures ($pO_2$). For the typical experimental synthesis and high temperature operating conditions, the NiO/Ga$_2$O$_3$ interface is chemically unstable based on this calculation. Instead, these conditions in the experimental range favor the formation of NiGa$_2$O$_4$ with spinel structure shown in the inset of Figure 3, supporting the observed spinel NiGa$_2$O$_4$ (or Ni-substituted γ-phase Ga$_2$O$_3$) found from TEM analysis of the NiO-Ga$_2$O$_3$ interface after thermal cycling.

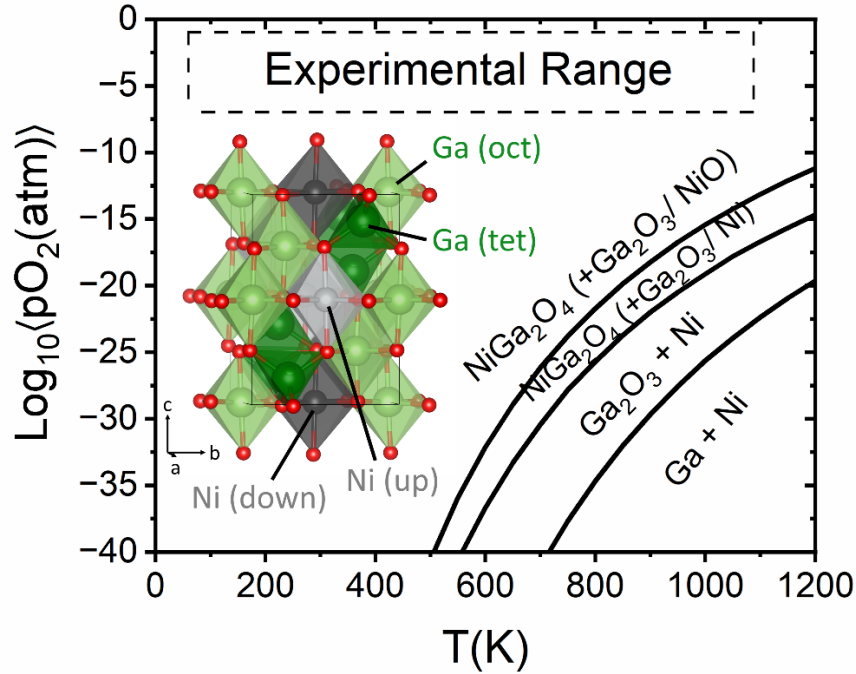

Figure 3: Predicted thermodynamic phase diagram of Ni-Ga-O system as a function of oxygen partial pressure (pO$_2$) and temperature. The "Experimental Range" box highlights the typical experimental synthesis and operation conditions for devices made from these materials. The parentheses in the figure indicate that depending on the synthesis condition, Ga$_2$NiO$_4$ can form as a single phase or coexist with either Ga$_2$O$_3$ or NiO/Ni. Inset: shows a Vesta[28] image of inverse spinel NiGa$_2$O$_4$ crystal structure (*P4$_1$22*). Darker and lighter colors indicate unique sites for Ga and Ni cations. "Up" and "down" refer to applied electron spins.

To better understand the potential impact of NiGa$_2$O$_4$ on the device performance as an interlayer in NiO-Ga$_2$O$_3$ heterojunction, we investigate the native charged point defect energetics of NiGa$_2$O$_4$ using density functional theory (DFT) calculations. Details of the calculations are described in the supplementary material. We model the ordered inverse spinel (*P4$_1$22*) NiGa$_2$O$_4$ crystal structure with anti-ferromagnetic spin since this is the lowest energy configuration according to our calculations. Its calculated *GW* band gap is ~ 3.3 eV. The formation energies of the lowest energy native defects are shown in Figure 4(a) for the most oxygen-rich synthesis condition (see Supplemental Table S1 for associated chemical potentials, $\Delta\mu$). The most relevant defects for *p*-type doping are those with low formation energy near the valence band edge ($E_F$=0 eV). $Ga_{Ni}$ substitutional is a shallow compensating donor, while the $Ni_{Ga}$ substitutionals on both the tetrahedral and octahedral sites are deep acceptors. The appearance of low energy octahedral cation anti-sites suggests $NiGa_2O_4$ will likely be a disordered inverse spinel (*Fd3m* space group) which is consistent with the findings of previous computational studies [29]. Due to its deep (0/-1) charge transition level, $Ni_{Ga}$ cannot act as an intrinsic dopant. However, Ni vacancies ($V_{Ni}$) can provide acceptors. Altogether, this sets an equilibrium Fermi level about 1.1 eV above the valence band edge, indicating that this material is weakly intrinsic *p*-type. In addition, there may be room to increase the number of holes beyond those provided from $V_{Ni}$ with an extrinsic acceptor dopant. While thin films of $NiGa_2O_4$ have seldom been reported, microstructured $NiGa_2O_4$ samples have been demonstrated as *p*-type materials useful for gas sensing, water splitting and energy storage applications[30–34], which is consistent with results of our theoretical calculations.

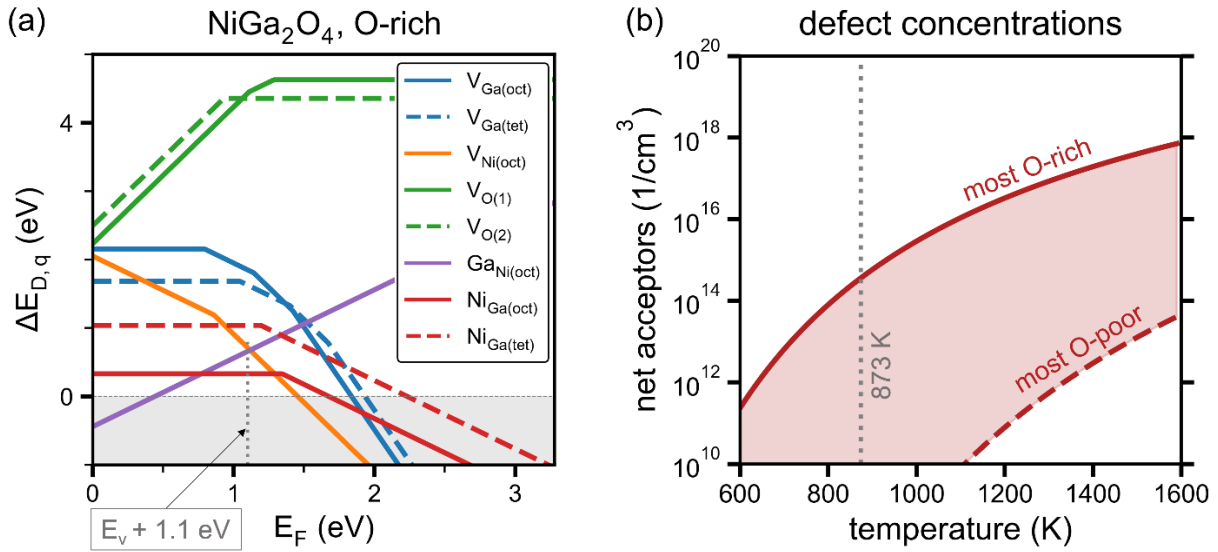

Figure 4. (a) Defect formation energy for lowest-energy native defects of inverse spinel $NiGa_2O_4$ under the most oxygen-rich condition plotted from VBM ($E_F$=0) to CBM ($E_F$= bandgap). Subscripts indicate different Wyckoff sites. Equilibrium Fermi level of 1.1 eV above VBM is indicated for a synthesis temperature of 873 K. (b) Native net acceptor concentrations [$Ni_{Ga(oct)}$ – 2 $V_{Ni(oct)}$] as a function of synthesis temperature and oxygen environment. Experimental synthesis temperature for the $NiGa_2O_4$ layer is indicated.

We also use thermodynamic modeling[35] to estimate the net acceptor concentration for a given temperature and chemical potentials, assuming a standard effective density of states for intrinsic carrier concentrations and an Arrhenius relationship for charged defects. Figure 4(b) shows the achievable net acceptor concentration for intrinsic $NiGa_2O_4$ within the bounds of its thermodynamic stability from decomposition. See Supplemental Table S1 for the chemical potentials associated with the most O-rich and most O-poor synthesis conditions. Considering thermodynamic equilibrium, we would expect net acceptor concentrations to range from below $10^{10}$ to $10^{18}$ cm$^{-3}$ depending on the synthesis conditions. For the conditions used in our experiments (T= 873K; $pO_2=10^{-3}$ Torr), we predict an equilibrium net acceptor concentration $\approx 10^{11}$-$10^{14}$ cm$^{-3}$ in our $NiGa_2O_4$ layer.

Next, we demonstrate vertical heterojunction NiO/$NiGa_2O_4$/n$^-$$\beta$-$Ga_2O_3$/n$^+$$Ga_2O_3$ diodes (Figure 5(a) employing thin $NiGa_2O_4$ thin films intended as a passivation layer at the NiO-$Ga_2O_3$ interface for high temperature applications. For these devices, $NiGa_2O_4$ and NiO layers were grown by pulsed laser deposition on ~5 μm lightly doped *n*-type $Ga_2O_3$ drift layer without breaking vacuum to facilitate a high quality NiO/$NiGa_2O_4$ hetero-interface. This was done to achieve a homogeneous $NiGa_2O_4$ layer of the same orientation and thickness throughout the device, rather than a randomly nucleated $NiGa_2O_4$ islands observed in TEM (Fig.2). Then devices were fabricated and characterized as described above. The room temperature semi-logarithmic current-voltage (J-V) curves for the fabricated heterojunction p-n diode are shown in Figure 5(a) for different values of the device diameter. The diode turn-on voltage ($V_{ON}$) defined at the forward current density of ~ 10 A.cm$^{-2}$ is 1.9V for the NiO/$NiGa_2O_4$/$Ga_2O_3$ diode and a diode rectification ($I_{ON}$/$I_{OFF}$) ratio ~ $10^{10}$($\pm$3V) is obtained for this diode. In contrast, NiO/$Ga_2O_3$ devices fabricated without the thin $NiGa_2O_4$ layer show a lower $V_{ON}$ of 1.7 V and similar $I_{ON}$/$I_{OFF}$ ratio. The minimum differential specific on-resistance, $R_{on,sp}$ obtained for the NiO/$NiGa_2O_4$/$Ga_2O_3$ devices is 0.04 Ω-cm$^2$ as shown in Figure S3 (Supplementary Material), which is higher compared to 0.03 Ω-cm$^2$ for the NiO/$Ga_2O_3$ device without the $NiGa_2O_4$ layer.

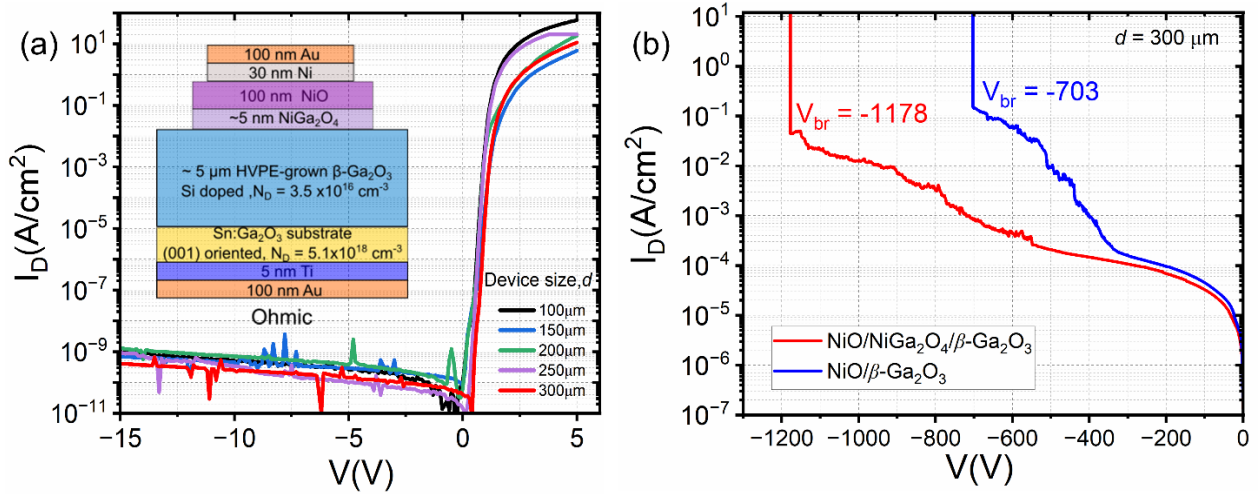

Figure 5(a) Semi-logarithmic J-V characteristics of the NiO/NiGa$_2$O$_4$/β-Ga$_2$O$_3$ HJ diode for different device diameters. Inset shows the schematic of the cross section for the fabricated NiO/NiGa$_2$O$_4$/β-Ga$_2$O$_3$ HJ diode (b) Reverse J-V characteristics of 300 μm diameter NiO/ NiGa$_2$O$_4$/β-Ga$_2$O$_3$ and NiO/β-Ga$_2$O$_3$ *p-n* heterojunctions diodes showing a breakdown voltage of ~1.18 kV and 703 V respectively.

Breakdown voltages ($V_{br}$) at which the devices show catastrophic failure are shown in the reverse J-V characteristics in Figure 5(b) for NiO/5 μm n$^-$β-Ga$_2$O$_3$/n$^+$Ga$_2$O$_3$ heterojunction devices with and without a thin NiGa$_2$O$_4$ layer. The breakdown voltage ($V_{br}$) at room temperature was measured with the samples fully immersed in FC-40 Fluorinert dielectric liquid using a Keysight B1505A Power device Analyzer with a reverse current limit of 1 mA. Measurement results show that the NiO/NiGa$_2$O$_4$/Ga$_2$O$_3$ devices have lower leakage current, and a higher breakdown voltage, $V_{br}$ =1.2 kV compared to 700 V for the NiO/Ga$_2$O$_3$ devices. This is attributed to the thin NiGa$_2$O$_4$ serving to provide improved edge termination resulting in suppressed electric field crowding in the device. Hence, NiGa$_2$O$_4$ is promising for a combined role of passivation layer for high temperature operation and voltage blocking layer for electric field management in NiO/NiGa$_2$O$_4$/Ga$_2$O$_3$ heterojunction *p-n* devices.

In summary, we showed NiGa$_2$O$_4$ as a *p*-type interfacial layer to improve the stability of the electrical properties in NiO/*β*-Ga$_2$O$_3$ heterojunction diode, as motivated by observed degradation in the electrical properties due to temperature induced interfacial instability. We showed the controlled thin film growth of the NiO/*β*-Ga$_2$O$_3$ reaction product (NiGa$_2$O$_4$) as an interfacial layer integrated in NiO/NiGa$_2$O$_4$/*β*-Ga$_2$O$_3$ heterojunction vertical power diodes contribute to the enhanced device reverse blocking capability. Non-field plated NiO/NiGa$_2$O$_4$/*β*-Ga$_2$O$_3$ heterojunction with I$_{ON}$/I$_{OFF}$ ratio and V$_{ON}$ of $10^{10}$ ($\pm$3 V) and 1.9V, respectively, supports a high breakdown voltage of 1.2kV for 300μm device compared to 700 V for the NiO/Ga$_2$O$_3$ devices, due to NiGa$_2$O$_4$ layer providing improved edge termination and reduced electric field crowding. Optimization of the conductivity and the carrier density of the NiGa$_2$O$_4$ layer through defect engineering and extrinsic doping should further improve device performance. The integration of thin NiGa$_2$O$_4$ layers between NiO and *β*-Ga$_2$O$_3$ is promising for realizing high performance *β*-Ga$_2$O$_3$ based vertical *p-n* devices with stable electrical performance at extreme environments such as at high temperatures.

## Supplementary Materials

See supplementary Materials for analysis of the NiO/Ga$_2$O$_3$ heterojunction device leakage current at -20V during thermal cycling, details of the thermodynamic and DFT calculations, and plot of the specific on-resistance of the NiO/NiGa$_2$O$_4$/Ga$_2$O$_3$ heterojunction diode.

## Author Declarations

The authors have no conflicts to declare.

## Data Availability

The data supporting the findings of this study are available within the paper and its Supplementary Material. Any additional data connected to the study is available from the corresponding author upon reasonable request.

**Acknowledgements**

This work was authored by the National Renewable Energy Laboratory (NREL), operated by Alliance for Sustainable Energy, LLC, for the U.S. Department of Energy (DOE) under Contract No. DE-AC36-08GO28308. Funding is provided by the Office of Energy Efficiency and Renewable Energy (EERE) Advanced Materials & Manufacturing Technologies Office. High-voltage breakdown measurements are supported by a laboratory directed research and development (LDRD) program at NREL. The views expressed in the article do not necessarily represent the views of the DOE or the U.S. Government.

**Author Contributions**

Kingsley Egbo – Conceptualization (equal), Data Curation (lead), Formal Analysis (lead), Methodology (lead), Software (equal), Validation (lead), Visualization (equal), Writing/Original Draft (lead).

Emily M. Garrity– Software (equal), Formal Analysis (supporting), Data Curation (supporting), Writing – review & editing (supporting).

William A. Callahan– Data curation (supporting), Resources (supporting), Writing – review & editing (supporting).

Chris Chae: Formal Analysis (supporting), Data curation (supporting), Resources (supporting).

Cheng-Wei Lee: Software (equal), Data Curation (supporting), Writing/Original Draft (supporting).

Brooks Tellekamp: Conceptualization (equal), Data Curation (supporting).

Jinwoo Hwang: Formal Analysis (supporting), Data curation (supporting), Resources (supporting)

Vladan Stevanovic: Conceptualization (equal), Supervision (equal).

Andriy Zakutayev: Project administration (lead), Supervision (equal), Funding acquisition (equal), Writing – review & editing (supporting).

**Supplementary Materials for**

**NiGa$_2$O$_4$ interfacial layers in NiO/Ga$_2$O$_3$ heterojunction diodes at high temperature**

Kingsley Egbo[1,*], Emily M. Garrity[2], William A. Callahan[1,2], Chris Chae[3], Cheng-Wei Lee[2], Brooks Tellekamp[1], Jinwoo Hwang[3], Vladan Stevanovic[2], Andriy Zakutayev[1,†]

[1] National Renewable Energy Laboratory, Golden, CO 80401, USA

[2] Colorado School of Mines, Golden, CO 80401, USA

[3] Department of Materials Science and Engineering, The Ohio State University, Columbus, Ohio 43210, USA

*Kingsley.Egbo@nrel.gov, †Andriy.Zakutayev@nrel.gov

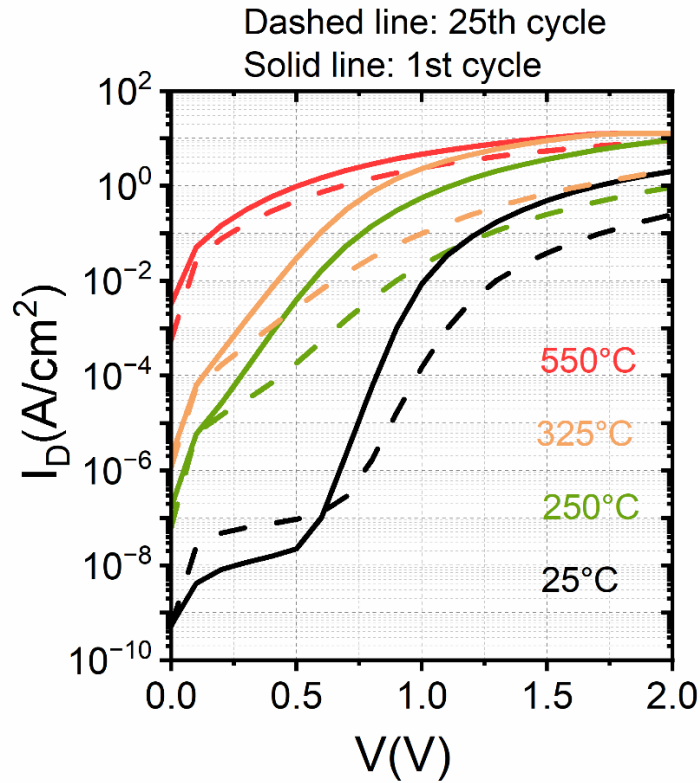

Figure S1: (a) Forward voltage J-V curve of the NiO/Ga$_2$O$_3$ *p-n* heterojunction device before and after the completion of 25 high temperature thermal cycles between 25-550°C. Solid and dashed lines are measured at the start of cycle 1 and end of cycle 25 respectively.

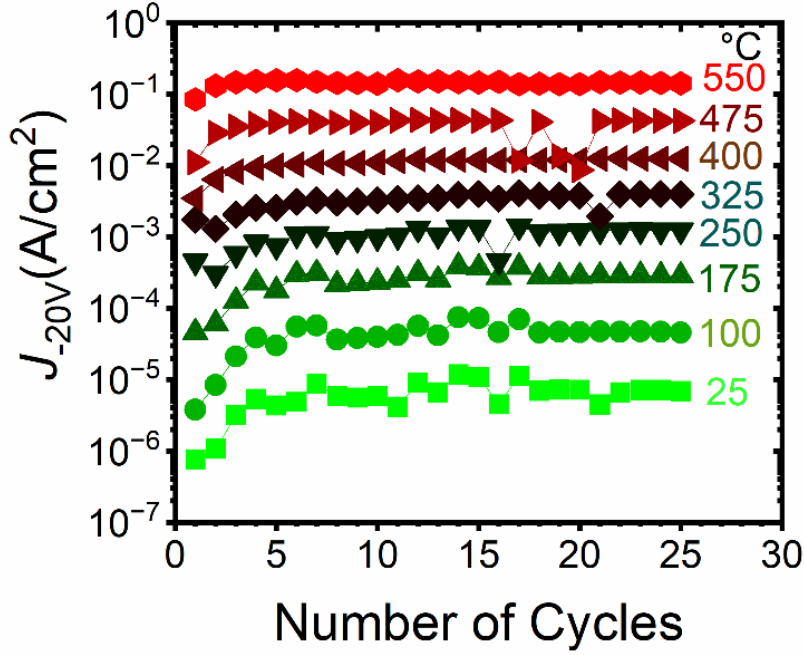

Figure S2: (a) Summary of the NiO/Ga$_2$O$_3$ *p-n* heterojunction device leakage current density evolution at – 20V across the 25 cycles for different temperatures.

**Details of the Thermodynamic Calculations:**

The Gibbs formation energy of a compound, $\Delta G_f(T)$, in units of eV/atom is calculated by,

$$\Delta G_f(T) = \Delta H_f(298.15K) + G^{\delta}_{SISSO}(T) - \sum_{i}^{N} \alpha_i\, G_{i,exp}(T) \tag{1}$$

where $\Delta H_f(298.15K)$ is the standard formation enthalpy of the compound and is calculated using DFT total energies from the NREL MatDB[1] and fitted elemental-phase reference energies[2]. $\alpha_i$ is the stoichiometric weight for element $i$ in the compound and we use experimental values from FactSage[3] for absolute Gibbs free energy of element $i$. $G^{\delta}_{SISSO}(T)$ (eV/atom) captures temperature dependency of the compound due to phonon and is defined as,

$$G^{\delta}_{SISSO}(T) = (-2.48 \times 10^{-4} \times \ln(V) - 8.94 \times 10^{-5} mV^{-1})T + 0.181 \times \ln(T) - 0.882$$

where $V$ is the atomic volume (Å$^3$/atom) of the compound and $m$ is the reduced atomic mass (amu).

Lastly, we collect $\Delta G_f(T)$ for all the competing phases and use convex hull analysis to calculate the grand potential phase diagram. Ideal gas law, e.g., $\Delta \mu_O = \frac{1}{2} k_B T ln(pO_2)$, is used to connect oxygen chemical potential ($\mu_O$) to oxygen partial pressure ($pO_2$), where $\Delta \mu_O$ is the deviation from the standard oxygen chemical potential ($\mu_O^°$) with $\mu_O = \mu_O^° + \Delta \mu_O$.[4]

**Details of DFT defect calculations:**

We follow the established supercell approach[5,6] to calculate the formation energy ($\Delta E_{D,q}$) for each defect $D$ (vacancies, anti-sites, and interstitials) in the charge state $q$ as a function of Fermi energy ($E_F$) at specific synthesis conditions which are described by elemental chemical potentials ($\mu_i$). We calculate this using.

$$\Delta E_{D,q} = [E_{D,q} - E_H] + \sum_i n_i \mu_i + qE_F + E_{corr}$$

where $E_{D,q}$ is the total energy of the supercell containing the defect, $E_H$ is the total energy of the host supercell with no defect or net charge, $n_i$ is the number of atoms of an element $i$ which is removed ($n_i > 0$) or added ($n_i < 0$) to form the defect, and $E_{corr}$ is the grouped correction term applied to alleviate various artifacts that arise from using periodic boundary conditions We use a 448-atom supercell of the ordered inverse spinel $NiGa_2O_4$ (tetragonal $P4_122$).. In this structure, all Ni atoms occupy the octahedral cation sites along with half of the Ga atoms and the other half of the Ga atoms are on the tetrahedral sites. To each Ni atom we apply a Hubbard $U$ correction of +3 eV [7,8] and spin polarization to create an anti-ferromagnetic spin configuration. We relax the defect supercells with density functional theory (DFT) generalized gradient approximation (GGA)[9,10] as implemented in the Vienna *ab*-initio simulation package[11] using a Gamma-only *k*-point mesh, a plane-wave cutoff energy of 340 eV, and a force convergence criteria of 5 meV/Å. We also perform self-consistent *GW* calculations with fixed wave functions[12,13] to adjust the band edges and correct the band-gap underestimation of GGA. The chemical potentials are calculated using the fitted elemental-phase reference energy approach [8]. A static dielectric constant of 10.83 is calculated using density functional perturbation theory (dfpt)[14,15] and is used for the image-charge correction.

For the thermodynamic modeling, Fermi level is calculated self-consistently to obey charge neutrality. In other words, the net sum of the charge carriers contributed by the dominant defects, $Ga_{Ni}^{+1}$ and $V_{Ni}^{-2}$, and the intrinsic carriers equals to zero. The intrinsic carrier concentration is calculated assuming a standard density of states with electron and hole effective masses of 0.034 and 3.448, respectively. The charged defect concentrations are calculated assuming an Arrhenius relationship and using the formation energies of each defect for the given Fermi level and synthesis condition. The ideal gas law is used to relate oxygen chemical potential to temperature and oxygen partial pressure[4].

**Table S1.** Phase equilibrium and associated chemical potentials ($\Delta\mu$) that define the synthesis conditions used in $P4_122$ $NiGa_2O_4$ defect calculations.

| Condition | Equilibrium phases | $\Delta\mu_O$ | $\Delta\mu_{Ni}$ | $\Delta\mu_{Ga}$ |
|---|---|---|---|---|
| **O-rich** | O, $Ga_2O_3$, $NiGa_2O_4$, | 0.000 | -2.678 | -5.405 |
| **O-poor** | Ni, $Ga_2O_3$, $NiGa_2O_4$ | -2.678 | 0.000 | -1.388 |

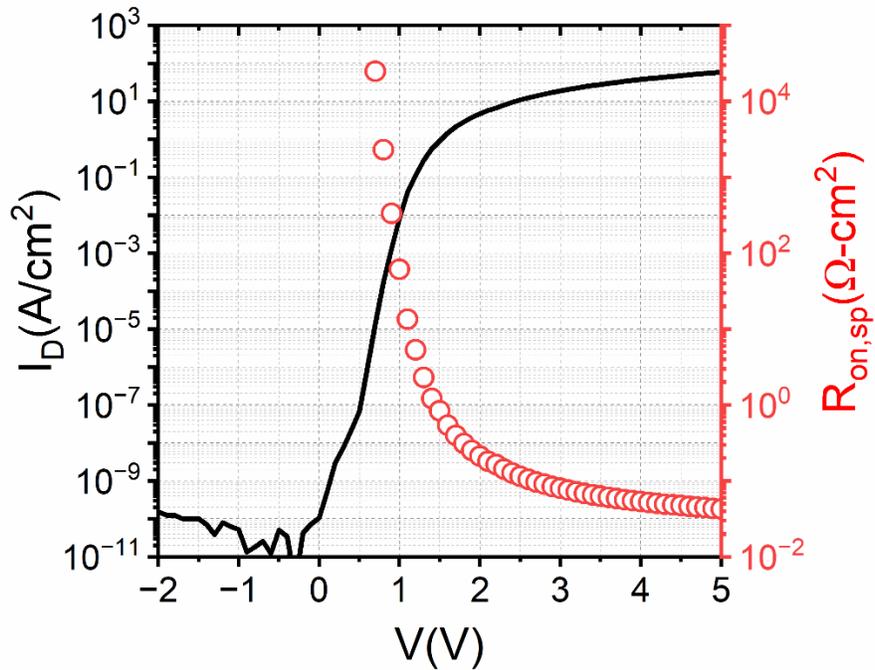

Figure S3: Forward J-V characteristics of the NiO/NiGa$_2$O$_4$/Ga$_2$O$_3$ heterojunction diode and the extracted differential specific on-resistance